\documentstyle[12pt,equation]{article}
\begin{document}
\newcommand{\cb}{\mbox{$\cos \! \beta$}}
\newcommand{\sib}{\mbox{$\sin \! \beta$}}
\newcommand{\cw}{\mbox{$\cos \! \theta_W$}}
\newcommand{\sw}{\mbox{$\sin \! \theta_W$}}
\newcommand{\aq}{\mbox{$a_{\tilde{q}_i}$}}
\newcommand{\bq}{\mbox{$b_{\tilde{q}_i}$}}
\newcommand{\asq}{\mbox{$a^2_{\tilde{q}_i}$}}
\newcommand{\bsq}{\mbox{$b^2_{\tilde{q}_i}$}}
\newcommand{\tanb}{\mbox{$\tan \! \beta$}}
\newcommand{\mx}{\mbox{$M_X$}}
\newcommand{\mc}{\mbox{$m_{\chi}$}}
\newcommand{\mcsq}{\mbox{$m^2_{\chi}$}}
\newcommand{\msq}{\mbox{$m_{\tilde{q}}$}}
\newcommand{\msqsq}{\mbox{$m^2_{\tilde{q}}$}}
\newcommand{\msqiv}{\mbox{$m^{-2}_{\tilde{q}}$}}
\newcommand{\msqf}{\mbox{$m^{-4}_{\tilde{q}}$}}
\newcommand{\be}{\begin{equation}}
\newcommand{\ee}{\end{equation}}
\newcommand{\een}{\end{subequations}}
\newcommand{\ben}{\begin{subequations}}
\newcommand{\beq}{\begin{eqalignno}}
\newcommand{\eeq}{\end{eqalignno}}
\renewcommand{\thefootnote}{\fnsymbol{footnote} }
\noindent
\begin{flushright}
MAD/PH/818\\
February 1994
\end{flushright}
\vspace{1.5cm}
\pagestyle{empty}
\begin{center}
{\Large \bf Supersymmetric Dark Matter: Relic Density and
Detection}\footnote{Talk presented at the International Conference on
Non--Accelerator Particle Physics, Bangalore, India, January 1994} \\
\vspace{5mm}
Manuel Drees\footnote{Heisenberg Fellow}\\
{\em Physics Department, University of Wisconsin, Madison, WI 53706, USA}
\end{center}

\begin{abstract}
The Lightest Supersymmetric Particle (LSP) makes a good Dark Matter (DM)
candidate, since its relic density quite naturally comes out close to the
cosmologically required value. This is true even in minimal Supergravity
models with radiative symmetry breaking, which have a rather small number of
free parameters. On the other hand, the experimental detection of SUSY DM
might be quite difficult, necessitating km$^2$ size neutrino detectors or
direct detection experiments with several tons of detector material.

\end{abstract}

\clearpage
\noindent
\setcounter{footnote}{0}
\pagestyle{plain}
\setcounter{page}{1}
\section*{1) Introduction}
The possibility that all known particles have superpartners \cite{1} with mass
$\sim 1$ TeV or less is now being taken quite seriously. The main motivation
for the introduction of supersymmetry is that it makes the theory technically
natural, i.e. protects the weak scale against large radiative corrections
\cite{2}. It has recently been found \cite{3} that supersymmetry also
facilitates the introduction of a Grand Unified gauge group at scale $\mx
\simeq 10^{16}$ GeV. Of more immediate interest for this talk is that in the
simplest, ``R--parity invariant'' SUSY models the lightest superparticle (LSP)
is absolutely stable. This means that some of the LSPs produced just after the
Big Bang should still be around today. Searches for exotic isotopes \cite{4}
put very stringent upper limits on the present abundance of charged and/or
strongly interacting particles; if its density is to be cosmologically
significant the LSP therefore has to be neutral.

Fortunately in many models the LSP more or less automatically comes out to
be the lightest neutralino, which satisfies this requirement. Assuming minimal
particle content the LSP is then a mixture of four different current
eigenstates: The bino $\tilde{B}$, the neutral wino $\tilde{W}$, and the
two higgsinos $\tilde{h}_1^0, \ \tilde{h}_2^0$. The mixing is governed by the
following mass matrix:
\be \label{e1}
{\cal M}^0 = \mbox{$ \left( \begin{array}{cccc}
M_1 & 0 & -M_Z \sw \cb & M_Z \sw \sib \nonumber \\
0 & M_2 & M_Z \cw \cb & -M_Z \cw \sib \nonumber \\
-M_Z \sw \cb & M_Z \cw \cb & 0 & -\mu \nonumber \\
M_Z \sw \sib & -M_Z \cw \sib & -\mu & 0 \end{array} \right). $} \ee
Here, $M_1$ and $M_2$ are SUSY breaking $U(1)$ and $SU(2)$ gaugino masses,
$\mu$ is the supersymmetric Higgs(ino) mass and $\tanb\equiv \langle H_2^0
\rangle / \langle H_1^0 \rangle$ is the ratio of vevs. For the remainder of
this talk I will assume gaugino masses to be unified at scale \mx, leading to
the following relation at the weak scale:
\be \label{e2}
M_1 = \frac {5}{3} \tan^2 \! \theta_W M_2 \simeq 0.5 M_2, \ee
at 1--loop order \cite{1}. This then implies that there are basically only
three possibilities for the LSP: \begin{itemize}
\item If $|\mu| \geq |M_2|$ the LSP is gaugino--like (mostly photino
for small $|M_1|$, mostly bino for $|M_1| \geq M_Z$); its mass is
$\sim M_1$.
\item If $|\mu| \simeq |M_1|$ the LSP is a mixed state, i.e. both higgsino
and gaugino components will be sizable.
\item If $|\mu| < |M_1|$ the LSP is higgsino--like with mass $\sim \mu$.
Notice that in this case the next--to--lightest sparticle, either the second
neutralino or the lighter chargino, will be close in mass to the LSP.
\end{itemize}
We will see in the next section that only one of these three types of
neutralinos makes a good particle DM candidate.

\section*{2) LSP relic abundance}
By now there are quite a few calculations or LSP relic abundances, stretching
back more than 10 years \cite{6}. There is general agreement that, over wide
regions of parameter space, the LSP does make a good particle DM candidate.
Here I will briefly summarize the results of ref.\cite{7}, which included
all 2--body final states accessible to LSP annihilation at tree level, and
which was among the first papers to use a spartcicle spectrum as predicted
in minimal supergravity models with radiative gauge symmetry breaking
\cite{8}.

In the very early Universe LSPs were in thermal equilibrium with ordinary
particles. However, as the Universe expanded and cooled the rate of reactions
involving LSPs eventually dropped below the expansion rate. At this point
the LSP density ``froze out", i.e. it basically remained constant (over
a co--moving volume). The larger the LSP annihilation cross section, the
smaller the freeze--out temperature and hence the smaller the LSP relic
abundance (due to the Boltzmann factor). Freeze--out approximately occurs at
$T_F \simeq \mc/20$, i.e. LSPs are nonrelativistic at freeze--out (hence
``cold DM"). It is therefore usually (but not always \cite{9}) sufficient to
expand the LSP annihilation cross section in powers of the relative cms
velocity $v$: \be \label{e3}
v \sigma_{\rm ann}(\chi \chi \rightarrow {\rm anything}) = a + b v^2
+ {\cal O}(v^4). \ee
The present relic density is then very roughly given by \be \label{e4}
\Omega_{\chi} h^2 \simeq \frac {0.08 {\rm pb}} {a + b/7}. \ee
Here $\Omega_{\chi}$ is the LSP relic density in units of the critical
(closure) density $\rho_c \simeq 2 \cdot 10^{-29}$ g/cm$^3$, and $h$ is the
Hubble expansion parameter in units of 100 km/(sec$\cdot$Mpc).

Note that all final states that are kinematically accessible contribute in
eq.(3). For each final state there are usually contributions from $t-$channel
diagrams (where a sfermion, neutralino or chargino is exchanged) as well as
$s-$channel diagrams (where a $Z$ or Higgs boson is exchanged).
$\Omega_{\chi}$ thus depends on the {\em entire} (s)particle spectrum. A more
or less exhaustive scan of parameter space is therefore only possible in
models where supersymmetry breaking is described by a small number of free
parameters, as compared to the dozens of parameters in general softly broken
supersymmetry. One particularly attractive class are minimal supergravity
(mSUGRA) models \cite{8}. Here one assumes one common nonsupersymmetric
squared scalar mass $m^2$ (sometimes called $m_0^2$) and one common gaugino
mass $M$ (sometimes called $m_{1/2}$), as well as one common trilinear soft
breaking parameter $A$. The sparticle spectrum is assumed to have this very
simple form only at ultrahigh energies near the Planck scale. Radiative
corrections, most conveniently described by a set of renormalization group
equations \cite{10}, change the spectrum at lower energies. In particular,
corrections involving the Yukawa coupling of the top quark can drive the
squared mass of one of the Higgs bosons to negative values, thereby
triggering electroweak gauge symmetry breaking.

In ref.\cite{7} we studied the LSP relic density in such mSUGRA
models\footnote{In this paper a relation between the SUSY breaking parameters
$A$ and $B$ was assumed, but relaxing this assumption does not alter the
overall conclusions.}. Fig.1, taken from that paper, shows contours of
$\Omega_{\chi} h^2$ in the $(M,\tanb)$ plane. We see that a good part of the
experimentally and theoretically allowed region (within the outer, dotted
lines) satisfies $0.25 \leq \Omega_{\chi} h^2 \leq 1$, i.e. allows for a flat
Universe as favored by inflationary scenarios for the range $0.5 \leq h \leq
1$ currently favored by cosmologists (area between the long dashed and solid
lines). The short dashed lines show locations of $s-$channel poles, i.e. $2
\mc = m_{Z,h,P}$, where the relic density is usually very small.

{}From this and similar figures one concludes that $\Omega_{\chi} h^2$ indeed
``naturally" comes out in the right ball park if $m$ and $M$ are of the order
of a few hundred GeV, just in the range expected from naturalness arguments.
Unfortunately the requirement $\Omega_{\chi} h^2 \leq 1$ does not translate
into a strict upper bound on sparticle masses, however. The reason is that
LSP annihilation can be greatly enhanced if $2 \mc$ is close to the mass
of the pseudoscalar boson $P$ (unlike $Z$ and $h$ exchange, $P$ exchange
gives an $S-$wave pole, i.e. contributes at order $v^0$). In mSUGRA this
requires large \tanb, since one needs a large $b$ Yukawa coupling to reduce
$m_P$. This loophole allows cosmologically viable SUSY breaking at scales well
above those favored by naturalness arguments.

Another important result is that a cosmologically interesting LSP is almost
always gaugino--like. This statement is true even in more general SUSY
models\cite{6}. Higgsino--like and mixed LSPs have large couplings to gauge
and Higgs bosons; their density therefore drops rapidly for $\mc > m_W$,
reaching interesting values again only for $\mc > 0.5$ TeV, which is already
uncomfortably heavy for the lightest superparticle if SUSY is to stabilize the
gauge hierarchy. If $\mc < m_W$ their density is suppressed by the proximity
of $s-$channel poles (to make this argument watertight one has to include
co--annihilation of the LSP with the next--to--lightest sparticle \cite{11}).
In mSUGRA $|\mu|$ has to be increased along with $m_t$ to give the proper $W$
and $Z$ boson masses, so that for $m_t \geq 150$ GeV the LSP is almost always
gaugino--like; this can be regarded as one of the successes of such models.

\section*{3) Search for LSP annihilation in the Sun and Earth}
How would one go about looking experimentally for the supersymmetric DM
particles that are predicted to surround us? One idea \cite{12} is to look for
energetic muon neutrinos emerging from the centre of the Earth or Sun. DM
particles have a finite chance to interact with nuclei in celestial bodies,
thereby losing enough energy to become trapped gravitationally. In subesquent
interactions those particles lose even more energy, and finally they
become concentrated in the centre of these bodies. After some time so many
LSP particles should have accumulated in the centre of the Sun and Earth that
they begin to annihilate at a significant rate; capture and annihilation
eventually reach equilibrium\footnote{In case of the Sun equilibrium should
have been reached long ago; for the Earth this is only true if LSPs are not
too heavy \cite{13}}. Some fraction of annihilation events will give rise
to energetic muon neutrinos, which can be detected in nucleon decay detectors
like Kamiokande \cite{13} or in dedicated neutrino detectors like AMANDA
\cite{14}. The total signal rate is proportional to \be \label{e5}
{\rm Signal} \propto ({\rm capture \ rate}) \cdot \sum_X
Br (\chi \chi \rightarrow X) \int_{E_{thr}}^{m_{\chi}} d E_{\nu}
(E_{\nu})^2 \frac {d \Gamma} {d E_{\nu}}. \ee
Since we assume dynamical equilibrium between capture and annihilation the
overall size of the signal can be expressed in terms of the capture rate.
However, not all final states $X$ contribute equally to the signal; rather,
it is proportional to the third moment of the neutrino energy spectrum
characteristic for that final state. (One factor of $E_{\nu}$ appears since
the $\nu \rightarrow \mu$ cross section grows with energy, the second is due
to the increased range of the produced $\mu$.) Present searches already
exclude \cite{13} some combinations of parameters; however, calculations
indicate \cite{15} that one needs a detector with effective area of at least
1 km$^2$ to cover most of the interesting parameter space.

Previous calculations \cite{12,15} only included final states $X$ accessible
at tree level. Especially for gaugino--like LSPs with mass below $m_t$ the
process \be \label{e6}
\chi \chi \rightarrow g g \ee
can also be relevant \cite{16}. The reason is that LSP annihilation even in
the Sun occurs almost at rest; hence only the ${\cal O}(v^0)$ term $a$ in the
annihilation cross section (\ref{e3}) is relevant. For $f \bar{f}$ final
states this term is proportional to $m_f^2$. LSPs at rest therefore
predominantly annihilate into the most massive accessible fermions ($c$ or
$b$ quarks or $\tau$ leptons). This is fortunate, since their (semi--)leptonic
decays can give rise to hard neutrinos. However, massless quarks do contribute
to the process (\ref{e6}) at the 1--loop level. The relative size of the
corresponding cross section is therefore roughly (for bino--like LSP)
\be \label{e7}
\frac {\sigma(\chi \chi \rightarrow gg)} {\sigma (\chi \chi \rightarrow
b \bar{b} )} \sim \left( \frac{\alpha_s}{\pi} \right)^2
\frac {\left( \sum_q Y_q^2 \right)^2} {Y_b^4} \left( \frac {m_{\chi}}
{m_b} \right)^2, \ee
which can exceed unity. Unfortunately gluons do not produce hard neutrinos;
the process (\ref{e6}) therefore reduces the signal by lowering the branching
ratio for detectable processes. Fig.2 shows that the reduction can amount to
a factor of 2 if LSPs are indeed gaugino--like and not much lighter than the
top quark. QCD corrections can therefore make the search for neutrino signals
from LSP annihilation significantly more difficult, but they should not make
it impossible.

\section*{4) Direct LSP detection}
Another possibility to test the existence of SUSY DM is to search for the
elastic scattering of relic LSPs off the nuclei in a detector \cite{17}.
In order to estimate the expected event rate one obviously needs to know
the LSP--nucleus scattering cross section. As a first step one has to
compute the $S$ matrix element for scattering off a single nucleon. In general
one distinguishes two kinds of LSP--nucleon interactions: Those that depend
on the spin of the nucleon and those that don't. The former interactions
arise from $Z$ boson and $\tilde{q}$ exchange while the latter are due to
scalar Higgs boson and $\tilde{q}$ exchange \cite{18}. This last contribution
-- the spin--independent $\tilde{q}$--exchange term -- involves some
subtleties, as shown in refs.\cite{19}. To see that we start from the
LSP--quark--squark interaction \be \label{e8}
{\cal L}_{\chi q \tilde{q}} = \bar{q} \sum_{i=1}^2 \left( \aq + \bq \gamma_5
\right) \chi \tilde{q}_i + h.c., \ee
where the couplings \aq, \bq\ depend on the mixing angles both in the
neutralino sector and between the two squark eigenstates of a given flavor
$\tilde{q}_i$. Eq.(\ref{e8}) allows to compute an effective LSP--quark
interaction by integrating out the squark fields; this is a good
approximation since one is interested in momentum transfers (much) less than
1 GeV, well below the squark mass. In leading order in inverse squark mass
one then gets a spin--independent interaction $\propto \asq-\bsq$, which
is nonzero only if chirality is broken in the (s)quark sector. In the Standard
Model and its supersymmetric extension chirality is broken only by terms
involving the quark mass, so that the leading (in powers of \msqiv)
spin--independent LSP--quark interaction is $\propto m_q$. (This is
obviously also true for the Higgs exchange contribution.)

This special role played by massive quarks was first recognized by Griest
\cite{18}. He and subsequent authors estimated the heavy quark contribution
to LSP--nucleon scattering using a trick due to Shifman et al.\cite{20}: By
closing the heavy quark line in a loop and attaching two gluons the matrix
element $\langle N | m_q \bar{q} q | N \rangle$ can (for $m_q \gg \Lambda_{\rm
QCD}$) be related to the matrix element $\langle N| F_{\mu \nu} F^{\mu \nu}|
N \rangle$ which in turn is related to the nucleon mass. Shifman et al. had
used this trick in an estimate of Higgs--nucleon scattering rates. However,
in the present case the LSP--quark interaction itself is due to the exchange
of a strongly interacting particle, a squark; by using the SVZ trick one
effectively contracts the $\tilde{q}$ propagator to a point {\em inside} a
loop, see fig.3. This procedure cannot be expected to produce reliable
answers when $m_q$ and $m_{\tilde q}$ are comparable (which might be true for
the top quark); it will also fail to reproduce higher order (in \msqiv)
terms.

We therefore computed \cite{19} the full LSP--gluon scattering amplitude.
Up to permutations of external lines there are four different Feynman
diagrams, only one of which is contained in the effective Lagrangian treatment
oulined above. This calculation reproduced the $(\asq-\bsq) F_{\mu \nu}
F^{\mu \nu} \bar{\chi} \chi$ term found previously, if $m_q^2 \ll \mcsq$; an
important new result was that this term is strongly suppressed if $m_q \geq
\mc$, which might well be the case for the top quark. In addition we found
a term $\propto (\asq+\bsq) \bar{\chi} \partial_{\mu} \gamma_{\nu} \chi
F^{\mu}_{\rho} F^{\nu \rho}$. When expanded in powers of \msqiv\ it only
starts at ${\cal O}(m^{-4}_{\tilde q})$; however, it is nonzero even if
chirality is conserved in the (s)quark sector, i.e. if $|\aq|=|\bq|$, and
therefore also receives contributions from light quarks. At first we were
unable to fully exploit this result, however, for two reasons. First, we did
not know the matrix element $\langle N| F^{\mu}_{\rho} F^{\nu \rho} | N
\rangle$. Second, the new contribution contained terms $\propto \log \frac
{\msqsq-\mcsq}{m_q^2}$, i.e. are infrared divergent for light quarks.

Following a tip by Ken--Ichi Hikasa we eventually realized \cite{19} that
these problems are actually related. The logarithms can be understood, and
re--summed to all orders, by using running parameters in an effective
LSP--quark interaction expanded to ${\cal O}(m^{-4}_{\tilde q})$; in
particular, quark operators mix with gluonic operators at one--loop level.
What is more, the resulting quark and gluon operators are nothing but a subset
of the so--called leading twist operators that appear in analyses of
deep--inelastic lepton--nucleon scattering. Once we had realized this our work
was basically done, since both the matrix of anomalous dimensions (necessary
to re--sum the large logarithms in case of light quarks) and the relevant
hadronic matrix elements are already well known (the latter from experiment;
nonperturbative effects related to long--distance QCD are absorbed in these
matrix elements, thereby solving the problem of IR divergencies). Our
calculation thus extends previous studies by including all orders in \msqiv\
for terms $\propto \asq-\bsq$ (which is important for $t$ quarks, as noted
above), and by including terms $\propto (\asq+\bsq) \left( \alpha_s \log \frac
{\msqsq} {m_q^2} \right)^n$ for all powers $n$. Comparing our results with
earlier papers we also identified several sign mistakes and missing factors of
2 in the literature.\footnote{For each term that had been treated in earlier
papers we found at least one reference that agreed with us, but usually also
at least one that did not.}

Using the basic LSP--nucleon cross section we computed LSP scattering rates in
a Ge detector using the standard formalism \cite{19}. Some results are shwon
in figs. 4 and 5, where we have again used (s)particle spectra as predicted by
mSUGRA. The results are quite sobering: Only in a small slice of parameter
space (for positive $M \leq 100$ GeV just outside the region excluded by LEP)
is the rate as large as 0.1 event/(kg day), which is the approximate limit of
sensitivity that the next round of experiments aims for. On the other hand,
even for quite modest sparticle masses of a few hundred GeV the rate can be as
small as $10^{-3}$ events/(kg day), see Fig.4, or even 10$^{-4}$ events/(kg
day) (Fig.5), due to destructive interference between the exchange of the
light and heavy neutral scalar Higgs bosons.
In order to detect a signal at
this level one would not only have to assemble several tons of detector
material, cooled to millikelvin temperatures; one would also have to suppress
backgrounds by another 2 or 3 orders of magnitude compared to the goal the
next round of Ge detectors is aiming for. I should add here that in these
mSUGRA models squark exchange, including the terms $\propto \asq+\bsq$, is
not very important, since they predict $\mc \leq \msq/6$ for squarks of the
first two generations.\footnote{Technically this inequality follows from the
assumption of unified gaugino masses at scale \mx, as well as the fact that
first generation squarks cannot be much lighter \cite{8} than gluinos, since
at one--loop level the gluino mass gives a positive contribution to the squark
mass.} For a heavy nucleus like Ge73 spin--dependent interactions are usually
also quite small; the total interaction is therefore dominated by Higgs
exchange.

\section*{5) Summary and Conclusions}
The lightest supersymmetric particle, or, more specifically, a gaugino--like
neutralino, makes an excellent particle DM candidate in that its relic
density more or less automatically comes out in the right ballpark even
in the very restrictive class of models known as minimal supergravity models.
Detection of these neutralinos might prove quite difficult, however. In order
to cover a significant fraction of the allowed model parameter space one
needs km$^2$ size neutrino detectors, or direct detection experiments that
are sensitive to one event per ton of detector and day. This conclusion is
somewhat more pessimistic than that of earlier studies, partly since
increasing lower bounds on sparticle and Higgs boson masses exclude models
with large LSP--nucleon scattering rates and partly because we insist that
the LSP should have a sizable relic density to be considered a viable DM
candidate. Previous studies often found large detection rates for
higgsino--like or mixed LSPs, assuming a fixed local density, and ignoring
the fact that such an LSP would tend to have a rather low overall relic
density. Moreover, at least for the heavy top quark now favored by
experiments, supergravity models predict the LSP to be gaugino--like.

One should also be aware that particle DM searches test not only particle
physics but also galaxy formation models, which are necessary to estimate the
density and velocity (hence the flux) of LSPs in the vicinity of the solar
system. Moreover, if a signal is detected it is not clear whether it can be
established as being due to superparticles (as opposed to, say, massive
neutrinos). For this reason it should be clear that DM searches can never
replace collider searches for supersymmetry; they might, however, allow us to
``see" for the first time the stuff that most of the Universe is made of.

\noindent
\subsection*{Acknowledgements}
I thank the organizers for inviting me to this very well planned and executed
meeting. I also wish to express my gratitude to Mihoko Nojiri, stern
disciplinarian and collaborator in the work summarized in this talk. This work
was supported in part by the U.S. Department of Energy under contract No.
DE-AC02-76ER00881, by the Wisconsin Research Committee with funds granted by
the Wisconsin Alumni Research Foundation, by the Texas National Research
Laboratory Commission under grant RGFY93--221, as well as by a grant from the
Deutsche Forschungsgemeinschaft under the Heisenberg program.

\clearpage
\section*{Figure Captions}

\renewcommand{\labelenumi}{Fig.\arabic{enumi}}
\begin{enumerate}

\item 
Example of contours of constant $\Omega_{\chi} h^2$ in the $(M,
\protect{\tanb})$ plane for $m_t=140$ GeV and $m=250$ GeV (1: solid lines;
0.25: long dashed). The region outside the outer, dotted lines is excluded by
various experimental and theoretical constraints other than the DM relic
density.

\item 
Reduction of the signal of energetic muon neutrinos from LSP
annihilation in the sun due to QCD effects. In the light (dark) shaded region
the signal is reduced by at least 10 (50) \%. The reduction is very small
once $\protect{\mc} > m_t = 150$ GeV.

\item   
The steps from LSP--quark scattering via Higgs and $\tilde{q}$
exchange (left) to an effective LSP--quark interaction (centre) and its
connection to the gluonic matrix element $\propto F_{\mu \nu} F^{\mu \nu}$
(right). Note that in the last step the squark propagator appears inside
the loop.

\item   
Contours of constant counting rate in a 76 Ge detector in mSUGRA. The
central region between the dotted lines is excluded experimentally, mostly due
to LEP chargino and Higgs searches. The solid line is the contour of
0.1 events/(kg day); in the narrow wedges between the dot--dashed lines
$\Omega_{\chi} h^2<0.05$.

\item  
The \protect{\tanb} dependence of the LSP counting rate in a 76 Ge
detector in mSUGRA. The minimum at \tanb=5 is due to destructive interference
between the two Higgs exchange diagrams.

\end{enumerate}
\end{document}